\documentclass[sigconf]{acmart}

\usepackage{graphicx}
\usepackage{booktabs}
\usepackage{subcaption}
\usepackage{algorithm}
\usepackage{algpseudocode}
\usepackage{amsmath}
\usepackage{tikz}
\usepackage{pgfplots}
\pgfplotsset{compat=1.18}
\usetikzlibrary{positioning,shapes,arrows}

\acmConference[Preprint]{Preprint}{May 2026}{USA}

\copyrightyear{2026}
\setcopyright{cc}
\setcctype{by-nc-nd}

\begin{document}

\title{Scaling Search Relevance: Augmenting App Store Ranking with LLM-Generated Judgments}

\author{Evangelia Christakopoulou}
\affiliation{%
  \institution{Apple}
  \city{Cupertino}
  \state{CA}
  \country{USA}
}
\email{echristakopoulou@apple.com}

\author{Vivekkumar Patel}
\affiliation{%
  \institution{Apple}
  \city{Cupertino}
  \state{CA}
  \country{USA}
}
\email{vpatel22@apple.com}

\author{Hemanth Velaga}
\affiliation{%
  \institution{Apple}
  \city{Seattle}
  \state{WA}
  \country{USA}
}
\email{h_velaga@apple.com}

\author{Sandip Gaikwad}
\affiliation{%
  \institution{Apple}
  \city{Cupertino}
  \state{CA}
  \country{USA}
}
\email{sandip_gaikwad@apple.com}

\author{Sean Suchter}
\affiliation{%
  \institution{Apple}
  \city{Cupertino}
  \state{CA}
  \country{USA}
}
\email{ssuchter@apple.com}

\author{Venkat Sundaranatha}
\affiliation{%
  \institution{Apple}
  \city{Cupertino}
  \state{CA}
  \country{USA}
}
\email{vsundaranatha@apple.com}

\begin{abstract}
Large-scale commercial search systems optimize for relevance to drive successful sessions that help users find what they are looking for. To maximize relevance, we leverage two complementary objectives: behavioral relevance (results users tend to click or download) and textual relevance (a result's semantic fit to the query). A persistent challenge is the scarcity of expert-provided textual relevance labels relative to abundant behavioral relevance labels. We first address this by systematically evaluating LLM configurations, finding that a specialized, fine-tuned model significantly outperforms a much larger pre-trained one in providing highly relevant labels. Using this model as a force multiplier for human annotation, we generate millions of textual relevance labels to overcome the data scarcity. We show that augmenting our production ranker with these textual relevance labels leads to a significant outward shift of the Pareto frontier: offline NDCG improves for behavioral relevance while simultaneously increasing for textual relevance. These offline gains were validated by a worldwide A/B test on the App Store ranker, which demonstrated a statistically significant $\textbf{+0.24\%}$ increase in conversion rate, with the most substantial performance gains occurring in tail queries, where the new textual relevance labels provide a robust signal in the absence of reliable behavioral relevance labels.

\end{abstract}


  
  
\maketitle


\section{Introduction}
\label{intro}

Search ranking systems for large-scale digital marketplaces like the App Store are critical for app discovery and user satisfaction. The effectiveness of these systems is paramount to ensuring users can easily discover and engage with the vast array of applications available. These systems need to optimize for both textual relevance (the semantic fit of a result, as judged by experts) and behavioral relevance (the likelihood of a user clicking or downloading a result) to ensure good user experience. While behavioral relevance labels are abundant, textual relevance labels generated by human judges are much rarer. This creates a fundamental problem: high-quality textual relevance labels are scarce and expensive to produce, creating a scalability bottleneck and leaving the textual relevance objective under-powered in multi-objective training.

To address this label scarcity, recent work has explored LLMs as scalable offline generators of training data~\cite{bonifacio2022inpars} and as automated judges~\cite{zheng2023judgingllmasajudgemtbenchchatbot}. In this work, we validate this ``LLM-as-a-Judge'' paradigm at industrial scale. We fine-tune an in-house LLM on existing judgments from human judges and use it to generate millions of new high-quality textual relevance labels across multiple storefronts and languages. We then use these new labels as an additional input to train the production ranker, effectively overcoming traditional annotation bottlenecks. This methodology does more than improve a single metric; it shifts outwards the behavioral-textual relevance Pareto frontier, enabling the ranker to achieve superior performance.

Our contributions are four-fold:

\begin{enumerate}
    \item We deploy large-scale experiments with multiple configurations through pretrained and fine-tuned in-house large language models, to determine the best way to get the highest quality relevance labels for the specific task of relevance annotation. 
    \item We generate millions of pointwise textual relevance labels across multiple languages, vastly expanding our  training data.
    \item We show that augmenting the training data of our ranker with millions of these LLM-generated labels leads to a significant outward shift of the behavioral-textual relevance Pareto frontier, strictly dominating the production model in offline NDCG metrics.
    \item We validate these gains through both offline evaluation and online A/B testing, confirming the effectiveness of our approach in a real-world commercial environment. Notably, the results reveal that the LLM-augmented model provides the greatest conversion rate improvements for tail queries, effectively bridging the relevance gap in areas where reliable behavioral data is most scarce.
\end{enumerate}

The rest of the paper presents the related work, details our approach
and presents a comprehensive analysis of our offline and online
results.
\section{Related work}
\label{rel_work}

\textbf{LLMs as Rankers.} Large language models (LLMs) have demonstrated success as zero-shot and few-shot rankers using pointwise, pairwise, and listwise prompting strategies~\cite{zhuang2023beyond, tourrank2024, zhuang2023setwise, mozafari2025good, pradeep2023rankzephyr, rankvicuna2023, sun2023rankgpt, llm4ranking2025, sun2024chatgptgoodsearchinvestigating}. 
Our work differs from these approaches as they mainly focus on deploying LLMs as real-time (re)rankers; instead, we use an LLM as an offline label generator to create large-scale textual relevance labels for a production ranker.

\textbf{LLMs as Judges.} The ``LLM-as-a-Judge'' paradigm utilizes LLMs as scalable offline annotators to mitigate the scarcity of labels from human judges~\cite{faggioli2023perspectives, zhang2024retrieve_annotate_evaluate, upadhyay2023ares, arabzadeh2025benchmarking, zheng2024semantic_eval, chiang2023can, gu2024survey, liu2023g, thomas2023llmsearcherprefs, macavaney2023oneshot}. 
Our paper distinguishes itself by using a fine-tuned, in-house LLM not just for evaluation, but as a way to generate millions of textual relevance labels used directly as training data for our production ranker. 
In contrast to approaches like RRADistill~\cite{choi2024rradistill}, which focuses on distilling the semantic understanding of large LLMs into smaller language models with specialized architectural modifications, our work utilizes pointwise labels to train the production ranker without modifying its internal architecture.

\textbf{Multi-Objective Learning to Rank (LTR).} Multi-objective learning to rank (LTR)~\cite{liu2009learning} is a principled way to balance diverse objectives, such as different forms of relevance or fairness~\cite{moo_ltr_distill_2024, meng2025generative, moo_recsys_genai_2025, querywise_fair_moo, momma2023multi, carmel2020multi}. Our work operates within this framework, but rather than introducing a new objective, we focus on strengthening the existing textual relevance objective by dramatically expanding its label coverage via an LLM. This differs from the approach of Liu
et al.~\cite{liu2024towards} who also combine content and behavioral signals but do so by applying a sigmoid transformation to the LLM-generated label, whereas we utilize a data-mixing approach within our multi-objective framework.

\textbf{Preference Alignment and Prompting for Ranking Objectives.}
A growing body of research examines how to align LLM preferences with ranking-oriented objectives~\cite{zhao2024opo_ndcg, modeling_rank_icl2025, raja2025aligning, yuan2026unifyingrankinggenerationquery}. 
The common goal of these methods is to improve the LLM's ability to act as a ranker itself. While our work is inspired by these alignment techniques in our prompt and model design, our goal is fundamentally different. We do not seek to deploy the LLM as a ranker, but to obtain relevance labels that align with the existing human-rated rubric and can be used for training the ranker.

\textbf{Positioning of Our Work.}
Our contribution connects these strands in a large-scale production system. We demonstrate that using a fine-tuned LLM as an offline judge to augment the training data for a multi-objective ranker is a scalable and effective industrial strategy. We show this approach shifts outwards the behavioral-textual relevance Pareto frontier, leading to measurable improvements in both offline metrics and online A/B tests.
\section{Proposed approach}
\label{methodology}

Our approach systematically augments the training data for our production ranker. It consists of two main stages: \begin{enumerate}
\item{generating millions of high-quality, textual relevance labels, and }
\item{integrating these new labels into our multi-objective ranker training pipeline.}
\end{enumerate}

\subsection{LLM relevance labels generation}
\label{sec:llm_judgments}

\begin{figure}[h!]
\centering
\fbox{%
\begin{minipage}{\dimexpr\linewidth-2\fboxsep-2\fboxrule}                                                    

\small
Imagine you are an App Store evaluator. You are given a user search term and an app returned for this query, with app metadata 1, app metadata 2, app metadata 3.

\par\medskip
The goal is to choose one of the following labels for the app given the query:
label\_1, label\_2, label\_3, label\_4, label\_5.

\par\medskip
Description of relevance levels is as follows: \ldots

\par\medskip
\textbf{Strict guideline:} Your response should only be the label and nothing else.

\par\medskip
\noindent\textbf{Example 1:} query: query\_1; app: app\_1 with associated app metadata: app\_metadata\_1, app\_metadata\_2, app\_metadata\_3; label: label\_1.

\noindent\textbf{Example 2:} query: query\_2; app: app\_2 with associated app metadata: app\_metadata\_1, app\_metadata\_2, app\_metadata\_3; label: label\_5.

\par\medskip
Now generate the label for this: query: query\_target; app: app\_target with associated app metadata: app\_metadata\_1, app\_metadata\_2, app\_metadata\_3.
\end{minipage}%
}
\caption{Example few-shot prompt for query (query\_target) and returned app (app\_target).}
\label{fig:prompt}
\Description{A text box showing the few-shot prompt template used for LLM-based relevance label generation.}   
\end{figure}

To address the scarcity of textual relevance labels from human judges, we adopt an LLM-as-a-Judge methodology where a model acts as a scalable offline annotator. We leverage historical search logs aggregated across queries, containing $n \times m$ candidate query-app pairs. For each pair, our goal is to generate a pointwise textual relevance label.

Our framework supports both large-scale pretrained and specialized fine-tuned in-house models. While fine-tuning on existing judgments from human judges allows a model to better internalize our specific ranking rubric, using high-capacity pretrained models remains a viable path. By performing this generation offline, we create a force multiplier for human annotation without the latency constraints of real-time LLM reranking.

To ensure the generated labels align with our human judges, we construct prompts using the same query and app metadata available to them. An example prompt is shown in Figure~\ref{fig:prompt}. Our approach explores both zero-shot configurations and few-shot prompting, which incorporates previously rated examples from human judges to guide the model's judgment. We also use the judgments they have provided as labels for finetuning. These generated labels are  represented as either string or numeric values and they correspond to ordinal relevance levels matching the rubric used by the human judges. They are then integrated into our multi-objective ranker to strengthen the textual relevance objective.

\subsection{Multi-objective ranker training}
Our production ranker is designed to optimize relevance. This is achieved within a multi-objective framework to jointly optimize for textual and behavioral relevance. 

To achieve this, we construct a unique training dataset from two distinct label sources: downloads and clicks from aggregated App Store search logs (for behavioral relevance) and explicit textual relevance judgments (from human judges and the LLM-generated relevance labels described in Section~\ref{sec:llm_judgments}). Crucially, the same query-app pair, represented by an identical feature vector, can appear in the training data multiple times—once with a behavioral relevance label and once with a textual relevance label.

Our ranker training is a practical application of Multi-Objective Optimization (MOO) designed to find optimal solutions along the Pareto frontier. We employ a common MOO technique known as scalarization, where we combine the two objectives into a single objective function by creating a weighted mix of the training data. The strict separation of label sources during comparison is crucial, as it ensures the gradients for each objective are computed independently. 

This approach allows us to control the relative influence of each objective by treating the data mixing ratio as a tunable hyperparameter. By sampling or limiting rows from each data source, we can construct datasets with any desired mix (e.g., $90-10$, $70-30$, $50-50$), enabling us to systematically train different models that correspond to different points on the Pareto frontier.

\section{Offline Experimental Evaluation}
We conduct a two-stage offline evaluation to validate our approach: first, we assess the quality of the LLM-generated labels, and second, we measure their impact on the production ranker.

\subsection{Dataset}
\label{sec:offline_dataset}
Our evaluation utilizes two primary data sources: \begin{enumerate}
    \item {a large-scale dataset of millions of query-app pairs from historical App Store session logs aggregated across queries over a large time window, and} 
    \item{ a much smaller dataset of historical textual relevance judgments on query-app pairs from our human judges. }
\end{enumerate}
The second, smaller dataset of human judgments is split into training and validation sets, used to fine-tune and evaluate the LLM judge respectively. We then use the LLM judge to perform inference on the first, large-scale aggregated query-app log data to generate new textual relevance labels. 

Then, we construct the data for training the ranker by combining the two primary data sources: historical aggregated logs (for behavioral relevance) and labels from human judges (for textual relevance), but also the new LLM-generated labels (for improved textual relevance). We also split this dataset into training and validation for the ranker.

\subsection{Offline results on the LLM relevance labels}
\label{sec:offline_llm}

\subsubsection{Experimental Setup}
We compared three LLM configurations:
\begin{enumerate}
    \item \textbf{Pretrained 3B}: A pretrained $3$-billion parameter in-house model.
    \item \textbf{Pretrained 30B}: A pretrained $30$-billion parameter in-house model.
    \item \textbf{Finetuned 3B}: The $3$B model, fine-tuned on the training set of human judgments dataset described in Section~\ref{sec:offline_dataset}. 
\end{enumerate}
We experimented with various prompt configurations (zero-shot vs. few-shot, string vs. numeric labels) using query and app metadata. To determine the best configuration, we evaluated each model's ability to predict the relevance label assigned by human judges on a held-out validation set. We report precision, recall, and F1 scores across all relevance classes.

\subsubsection{Results}
\begin{table}[h!]
\centering
\caption{Offline evaluation of LLM configurations. The finetuned model (\texttt{FT-3B}) most closely reproduces the labels from the human judges. Pretrained models are labeled \texttt{PT}.}
\label{tab:offline_results_en}
\begin{tabular}{lccc}
\toprule
\textbf{Model} & \textbf{Precision} & \textbf{Recall} & \textbf{F1} \\
\midrule
\texttt{PT-3B}  & 0.300 & 0.309 & 0.287 \\
\texttt{PT-30B} & 0.424 & 0.402 & 0.382 \\
\texttt{FT-3B}  & \textbf{0.802} & \textbf{0.798} & \textbf{0.800} \\
\bottomrule
\end{tabular}
\end{table}

Table~\ref{tab:offline_results_en} shows the results for the best performing model configurations on the validation set, measuring how closely each model reproduces the labels provided by the human judges across all levels.  

We can see that the finetuned model (\texttt{FT-3B}) outperforms the pretrained models in all cases. This is notable, as it is expected to outperform the pretrained model with the same number of parameters, but it also outperforms the pretrained model with ten times more parameters, showcasing the value of finetuning. This result carries significant industrial implications: fine-tuning a smaller, more efficient model proved more effective than using its much larger, pretrained counterpart, offering a clear path to production with lower computational and operational costs. 

 While fine-tuning the 30B model remains a compelling direction for future work, our findings confirmed that the finetuned 3B model provided a sufficiently strong and cost-effective solution.

For brevity, we omit detailed results on prompt design, but note that few-shot prompts with string labels performed best. The labels generated by this optimal \texttt{FT-3B} configuration are used to augment the ranker's training data, which we evaluate next.

\subsection{Offline Ranker Results}
\label{sec:offline_ranker}

\begin{table*}[ht!]
\centering
\caption{Offline Ranker Performance (NDCG). Augmenting the training data with LLM-generated labels (\texttt{llm-augmented}) improves relevance metrics compared to the production model (\texttt{prod}).}
\label{tab:gbdt_results}
\begin{tabular}{lcccccc}
\toprule
& \multicolumn{6}{c}{\textbf{Relevance}}\\
\cmidrule(l){2-7}
 & \multicolumn{3}{c}{\textbf{Textual}} & \multicolumn{3}{c}{\textbf{Behavioral}} \\
\cmidrule(r){2-4} \cmidrule(l){5-7}
\textbf{Model} & \textbf{NDCG@1} & \textbf{NDCG@3} & \textbf{NDCG@7} & \textbf{NDCG@1} & \textbf{NDCG@3} & \textbf{NDCG@7} \\
\midrule
\texttt{prod} & 0.867 & 0.803 & 0.760 & 0.646 & 0.479 & 0.403 \\
\texttt{llm-augmented} & \textbf{0.868} & \textbf{0.805} & \textbf{0.761} & \textbf{0.652} & \textbf{0.484} & \textbf{0.407} \\
\bottomrule
\end{tabular}
\end{table*}


Following the generation of LLM-based relevance labels, we conducted a series of large-scale experiments to measure their impact on the performance of the production ranker. The primary goal was to see the effect on textual and behavioral relevance as we augmented the training data.

\subsubsection{Experimental Setup}
We compare two models:
\begin{enumerate}
    \item{\textbf{\texttt{prod}}: The production model, trained using the behavioral relevance labels from the aggregated App Store logs and the limited set of textual relevance labels from human judges.}
    \item{\textbf{\texttt{llm-augmented}}: The \texttt{prod} model, with its training data further augmented by millions of LLM-generated textual relevance labels generated as described in Section~\ref{sec:offline_llm}.}
\end{enumerate}
We hold out a validation set to measure performance using the NDCG@k metric~\cite{jarvelin2002cumulated}. The relevance score for each item used in the NDCG calculation is defined differently for each objective: for textual relevance, it is the judgment provided from a human judge; for behavioral relevance, it is a score derived from aggregated user clicks and downloads.
Note that the judgments used for computing the NDCG@k for the textual relevance are disjoint from those used for fine-tuning and validation of the LLM relevance labels; these are distinct, non-overlapping sets. 
\subsubsection{Results}

The results of our offline evaluation are summarized in Table~\ref{tab:gbdt_results}, in terms of behavioral and textual relevance NDCG at ranks $1$, $3$ and $7$ of returned results.

These results confirm our central hypothesis that adding LLM-generated labels helps improve the textual relevance of the ranker. They additionally improve the behavioral relevance of the ranker, showcasing that the increase of the training data with LLM-generated labels not only increases the semantic relevance of the results to the query, but also increases the likelihood of users downloading or clicking the results. The results are consistent across all different values of $k$. This demonstrates a true Pareto improvement, shifting the Pareto frontier outwards, meaning the \texttt{prod} model's performance is strictly dominated. 

Further experimentation also revealed that by varying the proportion of LLM-generated labels, we could then move along this new, superior frontier, rather than shifting the frontier further outward.

\section{A/B test results}

\begin{table}[ht!]
\centering
\caption{A/B test results of the \texttt{llm-augmented} model in terms of percentage improvement over the \texttt{prod} model with respect to conversion rate. The improvement shown is statistically significant.}
\label{tab:ab}
\begin{tabular}{lc}
\toprule
\textbf{Model} & \textbf{Conversion rate} \\
\midrule
\texttt{prod} & - \\
\texttt{llm-augmented} & \textbf{+0.24\%} \\
\bottomrule
\end{tabular}
\end{table}


\begin{figure}[t]
\centering
\begin{tikzpicture}
\begin{axis}[
    width=\linewidth,
    height=6cm,
    xlabel={$\log_e$ Query frequency},
    ylabel={Conversion rate percentage difference},
    ymin=0,
    grid=both,
    thick,
    mark=*,
    tick label style={font=\small},
    label style={font=\small},
]
\addplot coordinates {
(15,0.0351204)
(14,0.0102348)
(13,0.0468)
(12,0.0351041)
(11,0.0682958)
(10,0.0529419)
(9,0.033805)
(8,0.038832)
(7,0.120015)
(6,0.157162)
(5,0.220029)
(4,0.355334)
(3,0.486329)
(2,0.593207)
(1,1.27318)
(0,3.74038)
};
\end{axis}
\end{tikzpicture}
\caption{Conversion rate percentage difference of \texttt{llm-augmented} vs \texttt{prod} model across query frequency buckets. The lower-numbered buckets correspond to lower frequency - more tail queries, while the higher-numbered buckets correspond to higher frequency-more head queries.}
\label{fig:query_frequency}
\Description{A line chart showing conversion rate difference between the LLM-augmented and production models across query frequency buckets, with the largest improvements occurring for tail queries at low frequency buckets.}  
\end{figure}

We validated our approach with a large-scale online A/B test that ran on worldwide traffic, comparing our \texttt{llm-augmented} model against the production model (\texttt{prod}). 

As shown in Table~\ref{tab:ab}, the \texttt{llm-augmented} model demonstrated a statistically significant $\textbf{+0.24\%}$ increase in our primary metric, conversion rate, defined as the proportion of search sessions with at least one app download. While this number may appear small, it is considered a significant improvement for a mature industrial ranker. Across storefronts worldwide, the \texttt{llm-augmented} model outperformed \texttt{prod} in conversion rate in $89\%$ of the storefronts.

To understand the source of these gains, we analyzed the conversion rate lift across query frequency buckets, from low-frequency (tail) queries to high-frequency (head) queries. The results, shown in Figure~\ref{fig:query_frequency}, reveal that the most substantial improvements occur in the tail. This is because tail queries, by definition, lack sufficient user traffic to generate reliable behavioral relevance signals. Our \texttt{llm-augmented} model excels here because the newly added textual relevance labels provide a robust and accurate signal where the behavioral signal is sparse or absent, effectively closing the relevance gap.

\section{Conclusion and future work}
By systematically evaluating both large-scale pretrained and specialized fine-tuned in-house models, we demonstrate that LLMs can serve as a powerful force multiplier for human judges, effectively shifting the Pareto frontier in a production ranking pipeline. 
Our findings, validated through rigorous offline evaluation and online A/B testing, confirm the industrial scalability of the LLM-as-a-Judge paradigm and  highlight that LLM-generated labels are most impactful for tail queries, where they provide a robust signal in the absence of reliable behavioral data. 
Our work provides a practical blueprint for other large-scale search systems to overcome relevance-label scarcity and systematically improve their ranking models by leveraging the power of modern LLMs as offline data generators.

In the future, we plan to experiment with additional fine-tuned models, as well as additional types of prompt creation, such as  pairwise and listwise configurations that will allow us to generate labels for pairs or lists of apps for a particular query. 

\begin{acks}                                                                                                              
We would like to thank Don Dini and Vivek Kanojiya for inspirational and helpful discussions in the early stages of this work. We are grateful to Ian Fischer and Aashir Gajjar for their engineering contributions and technical support.                          
\end{acks}       
  
\bibliographystyle{ACM-Reference-Format}
\balance
\bibliography{references_new}

\end{document}